\documentclass[conference]{IEEEtran}

\usepackage[margin=1in]{geometry}
\usepackage{booktabs}
\usepackage{multirow}
\usepackage{siunitx}
\usepackage{amsmath,amssymb}
\usepackage{graphicx}
\usepackage[hypertexnames=false]{hyperref}
\usepackage{microtype}
\usepackage{xcolor}
\usepackage{listings}
\usepackage{algorithm}
\usepackage{algpseudocode}
\usepackage{tikz}
\usepackage{pgfplots}
\pgfplotsset{compat=1.18}

\hypersetup{
  colorlinks=true,
  linkcolor=blue,
  citecolor=blue,
  urlcolor=blue
}

\begin{document}

\title{Verbatim Data Transcription Failures in LLM Code Generation:\\A State-Tracking Stress Test}

\author{
\IEEEauthorblockN{
Mohd Ariful Haque\IEEEauthorrefmark{2}, 
Kishor Datta Gupta\IEEEauthorrefmark{2},
Mohammad Ashiqur Rahman\IEEEauthorrefmark{1},
Roy George\IEEEauthorrefmark{2}
}
\IEEEauthorblockA{\IEEEauthorrefmark{1}Florida International University}
\IEEEauthorblockA{\IEEEauthorrefmark{2}Clark Atlanta University}
}

\maketitle

\begin{abstract}
Many real-world software tasks require exact transcription of provided data into code, such as cryptographic constants, protocol test vectors, allowlists, and calibration tables. These tasks are operationally sensitive because small omissions or alterations can remain silent while producing syntactically valid programs. This paper introduces a deliberately minimal transcription-to-code benchmark to isolate this reliability concern in LLM-based code generation. Given a list of high-precision decimal constants, a model must generate Python code that embeds the constants verbatim and performs a simple aggregate computation. We describe the prompting variants, evaluation protocol based on exact-string inclusion, and analysis framework used to characterize state-tracking and long-horizon generation failures. The benchmark is intended as a compact stress test that complements existing code-generation evaluations by focusing on data integrity rather than algorithmic reasoning.
\end{abstract}

\section{Introduction}

Large language models (LLMs) are now widely used as programming assistants, and their evaluation culture has largely emphasized problem-solving ability: synthesizing algorithms from natural language, passing unit tests, or solving curated benchmark tasks \cite{chen2021codex,austin2021mbpp}. In contrast, a substantial portion of real-world software development is far less intellectually demanding but far more operationally sensitive. Many practical tasks require the faithful transfer of existing information into code without modification. Security configuration lists, allowlists, cryptographic constants, protocol test vectors, and migration data are common examples. In these settings, correctness is defined by exactness and completeness rather than creativity or insight. Code that appears syntactically valid but silently omits or alters provided data can introduce failures that are difficult to detect and costly to diagnose, particularly in security- and safety-critical systems.

A simple use-case illustrates the risk. Consider a developer using an LLM to generate a Python module that embeds a large list of cryptographic nonces or sensor calibration offsets supplied by a trusted specification. The intended computation is trivial, such as summing the values or validating their range. If the generated code omits even a small subset of constants or duplicates others, the module may still execute without errors and pass superficial tests, while silently violating correctness assumptions. In a security context, such silent corruption can weaken integrity checks, invalidate cryptographic guarantees, or introduce subtle inconsistencies that are extremely difficult to trace back to the generation step.

Despite impressive advances in reasoning and code synthesis, state-of-the-art LLMs are not designed to guarantee faithful long-form transcription. Their training objective optimizes next-token prediction under distributional similarity rather than exact symbolic copying. As output length increases, the model must implicitly track which elements have already been emitted and which remain outstanding. This internal bookkeeping is encoded only implicitly in attention patterns and hidden activations, rather than through explicit counters or memory structures. As a result, generation quality often degrades gradually rather than failing loudly. Models may skip items, prematurely terminate, or drift into repetitive or templated output while still producing code that looks structurally correct. This behavior can be understood as a state-tracking or generative amnesia problem: during long, detail-sensitive generation, earlier constraints lose influence, and the model’s internal representation of task progress becomes unreliable. These failures are inherently probabilistic, meaning that repeated runs under the same prompt can yield different and inconsistent omissions.

This paper examines a deliberately minimal transcription-to-code task designed to isolate this failure mode. The task provides a list of high-precision numeric constants and asks the model to generate Python code that embeds the numbers and computes a simple aggregate. The computation itself is trivial and easily verifiable; any error arises almost entirely from imperfect data transfer between the prompt and the generated code. This makes the task an effective stress test for LLM reliability in scenarios that closely resemble real-world security-sensitive software workflows. Our findings reveal a sharp scaling boundary: while some models reliably handle short lists, none of the evaluated state-of-the-art models consistently produce complete transcriptions for longer inputs. For large lists, many outputs contain none of the expected values, illustrating a dangerous combination of partial correctness and silent failure. Understanding and measuring this limitation is essential before LLMs can be trusted as autonomous components in secure software pipelines, where missing or corrupted constants can directly translate into security or safety risk.

\section{State tracking and ``amnesia'' in autoregressive generation}

The phrase ``LLM amnesia'' is not literal memory loss.
It is a convenient name for a structural mismatch between what the task demands and how an LLM produces text.
A typical LLM generates tokens one by one, conditioned on a limited context window of previous tokens \cite{vaswani2017attention}.
In many tasks, this is enough, because the output can be semantically coherent even if some low-level details drift.
In a transcription task, the low-level details \emph{are} the task.

The core difficulty is \emph{state tracking}.
When a model is asked to emit $N$ items in order, it must implicitly represent and update an internal ``cursor'' that corresponds to which items have already been emitted and which remain.
Unlike a traditional program, an LLM does not have an explicit loop counter or a protected data structure that guarantees progress.
Its state is a distributed pattern over activations that must be reconstructed at every generation step from the current context.
As the output grows, the context becomes longer, the signal for ``where we are'' is diluted among many similar tokens, and the model is prone to skipping items, repeating patterns, or abandoning the required format.

High-precision decimals are particularly punishing because they have low redundancy.
Natural language provides abundant constraints: grammar, semantics, and world knowledge.
A list of quasi-random decimals provides almost none.
If one digit is wrong, nearby digits do not ``pull'' the model back toward correctness.
From an information perspective, the task is closer to copying random bits than to paraphrasing a sentence, and the probability of a perfect run drops quickly as the required output length increases.
Even under an optimistic independence model where each item is copied correctly with probability $q$, the probability of a perfect transcription scales as $q^N$, which decays exponentially in $N$.
Our results indicate something stronger than exponential decay from a fixed $q$: the effective per-item fidelity itself degrades with $N$, producing a distinct accuracy cliff.

This makes transcription a useful stress tool.
It forces the model to sustain a brittle format for a long time; it is easy to validate, and it exposes error modes that can remain hidden on algorithmic benchmarks where tests cover only functional behavior and not the integrity of embedded data.

\section{Experimental setup}
We provide a prompt to an LLM that asks it to generate a Python script. In each run, the model is given an input text file containing N decimal values, one per line. The prompt instructs the model to copy these values exactly and assign them to variables in the generated code.

The output must be Python code only. The code is required to embed the provided numbers directly in source form (for example, as constants or as elements of a list) and compute their sum. The evaluation harness does not accept a precomputed numeric answer; it accepts only the generated code.

The notebooks supplied with the experiment show a workflow that loads the number file, inserts the list into the prompt, calls a model endpoint, and stores outputs for later validation.
The accompanying analysis spreadsheet documents the locally served model inventory and associates each model with a target input file.

We evaluate the 11 models listed in Table~\ref{tab:models}.
All quantitative results are computed from the provided validation reports, which contain per-run match statistics.
The experiment was executed in two independent batches.
We refer to them as Batch~A and Batch~B to keep the discussion date-free while preserving the independence of the runs.

\begin{table}[t]
\centering
\small
\begin{tabular}{ll}
\toprule
\textbf{Model (serving tag)} & \textbf{Label used in reports} \\
\midrule
\texttt{mistral-large:123b} & \texttt{mistral\_123B}\\
\texttt{deepseek-coder-v2:236b} & \texttt{deepseek\_236b}\\
\texttt{gpt-oss:20b} & \texttt{gpt-oss\_20b}\\
\texttt{llama3.3:70b} & \texttt{llama3.3\_70b}\\
\texttt{llama3.2:3b} & \texttt{llama3.2\_3b}\\
\texttt{gpt-oss:120b} & \texttt{gpt-oss\_120b}\\
\texttt{qwen3-coder:30b} & \texttt{qwen3-coder\_30b}\\
\texttt{deepseek-r1:70b} & \texttt{deepseek-r1\_70b}\\
\texttt{codestral:22b} & \texttt{codestral\_22b}\\
\texttt{codegemma:7b} & \texttt{codegemma7\_7b}\\
\texttt{codellama:34b} & \texttt{codellama34\_34b}\\
\bottomrule
\end{tabular}
\caption{Models evaluated in our transcription-to-code benchmark. The report labels are treated as identifiers for analysis; we do not assume architectural equivalence beyond the serving tags.}
\label{tab:models}
\end{table}

\begin{table*}[t]
\centering
\small
\begin{tabular}{lrrrrrr}
\toprule
\textbf{Batch} & \textbf{$N$} & \textbf{\#Models} & \textbf{\#Runs} & \textbf{Perfect runs} & \textbf{Perfect (\%)} & \textbf{Best match (\%)} \\
\midrule
Batch A & 100 & 11 & 1100 & 485 & 44.09 & 100.00 \\
Batch A & 300 & 11 & 1100 & 0 & 0.00 & 79.33 \\
Batch A & 500 & 11 & 1100 & 0 & 0.00 & 53.40 \\
Batch B & 100 & 11 & 1100 & 424 & 38.55 & 100.00 \\
Batch B & 300 & 11 & 1100 & 0 & 0.00 & 89.67 \\
Batch B & 500 & 11 & 1100 & 0 & 0.00 & 54.20 \\
\bottomrule
\end{tabular}
\caption{Summary of the two independent evaluation batches derived from the validation reports. ``Perfect'' means all expected numbers appear in the generated code as exact strings.}
\label{tab:batches}
\end{table*}

\subsection{Prompt for LLM}
We designed two different sets of instructions to test how well the models could handle the data. We call these Batch A and Batch B. In both cases, our main goal was to see if the model could take a list of input numbers and write Python code that includes those exact numbers, without making mistakes or leaving any out.

For \textbf{Batch A}, we formatted the input numbers as specific Decimal objects. We asked the model to generate Python code that declares a unique variable for every number in the list. The prompt explicitly told the model to output only the variable declarations and no other text. e.g.,

``Generate Python code that declares variables for each of the following decimal numbers (comma-separated): [Decimal('56745205.12641613782888039275'), Decimal('56116640.68338142986389638356'), Decimal('88248639.84894447969617237160'), Decimal('14180861.47871217335334137477'), Decimal('70279686.26509721828930332688'), Decimal('72342094.35080483565175368810'), Decimal('88035699.52161426017212064732')]. Each number must be assigned to its own variable in the returned code. Return only the Python variable declarations. Total variables to declare: 7."

For \textbf{Batch B}, we gave the model the numbers as a simple list separated by commas. We asked the model to write code that sums these numbers up. However, we included a strict rule: the model was not allowed to use lists, arrays, or dictionaries. Instead, it had to create a constant variable for each individual number. This forced the model to write out every number in the code explicitly. e.g.

``Generate Python code that sums the following numbers: 56745205.12641613782888039275, 56116640.68338142986389638356, 88248639.84894447969617237160, 14180861.47871217335334137477, 70279686.26509721828930332688, 72342094.35080483565175368810, 88035699.52161426017212064732. You must create constant variables for each number (no list or array or dict) and then sum them up. do not include any other text or comments in the code. Return python code not result."

\subsection{Validation metric}

The validation reports compute whether the generated code contains each expected numeric string as a substring.
Let $E=\{e_1,\ldots,e_N\}$ be the expected numbers as exact strings, and let $y$ be the generated output text.
We compute{\small
\[
\text{found\_count} = \sum_{i=1}^N \mathbb{1}[e_i \subset y],
\qquad
\text{match\_rate} = \frac{\text{found\_count}}{N}.
\]}
A run is \texttt{VALID} if and only if $\text{found\_count}=N$.
Otherwise it is \texttt{INVALID}.
Algorithm~\ref{alg:validator} summarizes this logic.

\begin{algorithm}[t]
\caption{Validator used to compute report fields (high-level pseudocode).}
\label{alg:validator}
\begin{algorithmic}[1]
\Require expected numbers as strings $E=\{e_1,\dots,e_N\}$, model output text $y$
\State $\text{found\_count} \gets 0$
\For{$i \gets 1$ to $N$}
  \If{$e_i$ is a substring of $y$}
    \State $\text{found\_count} \gets \text{found\_count} + 1$
  \EndIf
\EndFor
\State $\text{match\_rate} \gets \text{found\_count}/N$
\State \Return \texttt{VALID} iff $\text{found\_count}=N$ else \texttt{INVALID}
\end{algorithmic}
\end{algorithm}

This metric deliberately measures verbatim inclusion rather than semantic numeric equivalence.
If a model rewrites a decimal in scientific notation or rounds digits, the validator treats the value as missing even when it is numerically close.
That strictness is a feature for our purposes: the experiment targets the integrity of data embedding, not floating-point arithmetic.

\section{Results}

Table~\ref{tab:mainresults} aggregates performance across both batches.
For $N=100$, several models achieve near-perfect mean match rates and high perfect-run rates.
For example, \texttt{gpt-oss\_120b} produces perfect transcriptions in all recorded runs at $N=100$.
At $N=300$ and $N=500$, the situation changes qualitatively: across all models and both batches, the number of perfect runs is zero.

\begin{table*}[t]
\centering
\caption{Exact-string copy fidelity aggregated across both batches. ``Mean'' is the average match rate (fraction of expected numbers found). ``Perfect'' is the fraction of runs with all numbers present. ``Best'' is the maximum match rate observed for that model and $N$.}
\label{tab:mainresults}
\small
\begin{tabular}{lrrrrrr}
\toprule
\textbf{Model} & \multicolumn{2}{c}{\textbf{$N=100$}} & \multicolumn{2}{c}{\textbf{$N=300$}} & \multicolumn{2}{c}{\textbf{$N=500$}} \\
\cmidrule(lr){2-3}\cmidrule(lr){4-5}\cmidrule(lr){6-7}
 & Mean (\%) & Perfect (\%) & Mean (\%) & Best (\%) & Mean (\%) & Best (\%) \\
\midrule
\texttt{mistral\_123b} & 36.38 & 0.00 & 13.40 & 20.67 & 4.78 & 12.20\\
\texttt{deepseek\_236b} & 44.79 & 0.00 & 3.71 & 20.33 & 2.63 & 12.20\\
\texttt{gpt-oss\_20b} & 99.85 & 99.01 & 52.86 & 88.33 & 25.88 & 51.80\\
\texttt{llama3.3\_70b} & 99.67 & 66.50 & 23.22 & 48.00 & 1.47 & 27.00\\
\texttt{llama3.2\_3b} & 99.69 & 94.50 & 11.33 & 50.00 & 1.00 & 19.20\\
\texttt{gpt-oss\_120b} & 100.00 & 100.00 & 73.90 & 89.67 & 35.87 & 54.20\\
\texttt{qwen3-coder\_30b} & 43.95 & 0.00 & 0.50 & 18.67 & 5.09 & 12.00\\
\texttt{deepseek-r1\_70b} & 97.38 & 94.50 & 3.68 & 40.33 & 0.50 & 18.80\\
\texttt{codestral\_22b} & 9.39 & 0.00 & 0.56 & 5.00 & 0.45 & 11.80\\
\texttt{codegemma\_7b} & 29.67 & 0.00 & 1.20 & 11.33 & 0.34 & 4.40\\
\texttt{codellama\_34b} & 36.77 & 0.00 & 7.64 & 79.33 & 0.30 & 10.80\\
\bottomrule
\end{tabular}
\end{table*}

The aggregate behavior across all models shows a clear scaling collapse.
The mean match rate (averaged over all models) falls from 63.46\% at $N=100$ to 15.65\% at $N=300$ and 7.12\% at $N=500$.
Perfect runs fall from a substantial fraction at $N=100$ to zero at longer lengths, despite thousands of attempts.

A second pattern is visible in the tail behavior.
Table~\ref{tab:tails} reports median match rates and the proportion of ``zero-match'' runs, where none of the expected numeric strings appear in the output.
Several models that are strong at $N=100$ nevertheless exhibit a median of 0\% at $N=500$, meaning that more than half of their runs contain none of the expected numbers.
Even the strongest model in this dataset shows instability at $N=500$, with a wide spread between its lower and upper quantiles (analyzed below).

\begin{table*}[t]
\centering
\small
\begin{tabular}{lrrrr}
\toprule
\textbf{Model} & \textbf{$N=300$ Median (\%)} & \textbf{$N=300$ Zero (\%)} & \textbf{$N=500$ Median (\%)} & \textbf{$N=500$ Zero (\%)} \\
\midrule
\texttt{mistral\_123b} & 17.00 & 25.00 & 2.80 & 45.00\\
\texttt{deepseek\_236b} & 0.00 & 51.50 & 0.00 & 63.50\\
\texttt{gpt-oss\_20b} & 71.33 & 25.00 & 26.80 & 6.50\\
\texttt{llama3.3\_70b} & 23.33 & 34.50 & 0.00 & 87.00\\
\texttt{llama3.2\_3b} & 9.50 & 47.00 & 0.00 & 83.00\\
\texttt{gpt-oss\_120b} & 73.00 & 2.00 & 41.00 & 2.00\\
\texttt{qwen3-coder\_30b} & 0.00 & 95.50 & 0.00 & 52.00\\
\texttt{deepseek-r1\_70b} & 0.00 & 61.00 & 0.00 & 80.00\\
\texttt{codestral\_22b} & 0.33 & 41.50 & 0.40 & 27.00\\
\texttt{codegemma\_7b} & 0.00 & 50.50 & 0.00 & 88.00\\
\texttt{codellama\_34b} & 0.00 & 53.50 & 0.00 & 87.50\\
\bottomrule
\end{tabular}
\caption{Distributional indicators for long lists. ``Zero'' is the percentage of runs with \texttt{found\_count}$=0$. Medians and zero rates expose bimodality that mean rates can hide.}
\label{tab:tails}
\end{table*}

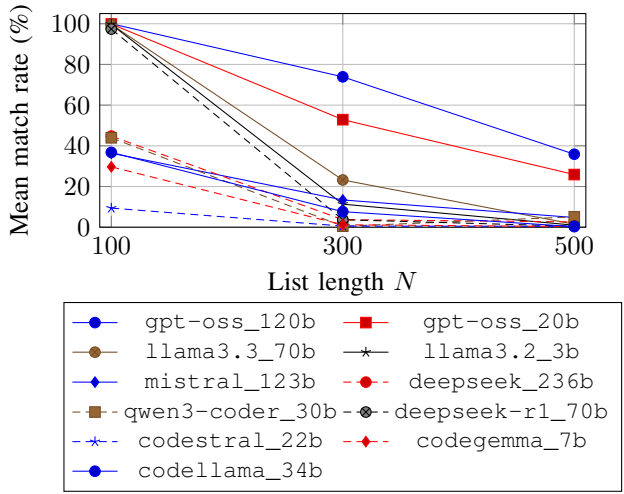
\begin{figure}[t]
\centering
\begin{tikzpicture}
\begin{axis}[
    width=\linewidth,
    height=0.55\linewidth,
    xlabel={List length $N$},
    ylabel={Mean match rate (\%)},
    xmin=90, xmax=510,
    ymin=0, ymax=105,
    xtick={100,300,500},
    ytick={0,20,40,60,80,100},
    grid=both,
    legend style={font=\small, at={(0.5,-0.35)}, anchor=north, legend columns=2},
]
\addplot+ coordinates {(100,100.00) (300,73.90) (500,35.87)};
\addlegendentry{\texttt{gpt-oss\_120b}}
\addplot+ coordinates {(100,99.85) (300,52.86) (500,25.88)};
\addlegendentry{\texttt{gpt-oss\_20b}}
\addplot+ coordinates {(100,99.67) (300,23.22) (500,1.47)};
\addlegendentry{\texttt{llama3.3\_70b}}
\addplot+ coordinates {(100,99.69) (300,11.33) (500,1.00)};
\addlegendentry{\texttt{llama3.2\_3b}}
\addplot+ coordinates {(100,36.38) (300,13.40) (500,4.78)};
\addlegendentry{\texttt{mistral\_123b}}
\addplot+ coordinates {(100,44.79) (300,3.71) (500,2.63)};
\addlegendentry{\texttt{deepseek\_236b}}
\addplot+ coordinates {(100,43.95) (300,0.50) (500,5.09)};
\addlegendentry{\texttt{qwen3-coder\_30b}}
\addplot+ coordinates {(100,97.38) (300,3.68) (500,0.50)};
\addlegendentry{\texttt{deepseek-r1\_70b}}
\addplot+ coordinates {(100,9.39) (300,0.56) (500,0.45)};
\addlegendentry{\texttt{codestral\_22b}}
\addplot+ coordinates {(100,29.67) (300,1.20) (500,0.34)};
\addlegendentry{\texttt{codegemma\_7b}}
\addplot+ coordinates {(100,36.77) (300,7.64) (500,0.30)};
\addlegendentry{\texttt{codellama\_34b}}
\end{axis}
\end{tikzpicture}
\caption{Mean copy fidelity vs.\ list length for a representative subset of models. Short-list performance is not predictive of long-list performance.}
\label{fig:scaling}
\end{figure}

\section{Analysis of failure modes}

The results support two distinct failure regimes that become prominent as $N$ increases.

The first regime is \emph{capacity-limited partial transcription}, where the output contains a substantial prefix of the expected numbers but stops short of completeness.
This is visible in the ``best'' scores at long lengths.
At $N=500$, the best observed runs for the two strongest models include roughly 260--270 numbers (\texttt{gpt-oss\_120b} reaches 271/500; \texttt{gpt-oss\_20b} reaches 259/500).
The clustering of maxima well below 500 is consistent with a hard ceiling such as an output-length constraint or an internal tendency to terminate after producing a long repetitive structure.
Without the raw output lengths, we cannot prove truncation, but the saturation pattern is difficult to explain purely as random per-number errors.

The second regime is \emph{derailment}, where the model produces code-like text that contains none of the expected numbers.
Derailment becomes common for long lists: 45.19\% of all $N=300$ runs and 56.50\% of all $N=500$ runs have \texttt{found\_count}$=0$.
Some models show a heavy-tailed mixture of these regimes, producing either a near-complete transcription or an almost total failure with little middle ground.
For example, at $N=300$ the median run for \texttt{gpt-oss\_20b} includes 214/300 numbers, yet 25\% of its runs contain zero expected numbers.
This is a textbook state-tracking instability: the model can follow the format, until it cannot.

Even when a model avoids derailment, it may still be unreliable.
At $N=500$, \texttt{gpt-oss\_120b} has only 2\% zero-match runs, but its lower-tail performance is poor: the 10th percentile includes about 65/500 numbers, while the median includes 205/500.
That spread matters for engineering because a single bad run can silently corrupt a downstream artifact.

\begin{figure}[t]
\centering
\begin{tikzpicture}
\begin{axis}[
    width=\linewidth,
    height=0.55\linewidth,
    xlabel={Percentile of runs (\%)},
    ylabel={Match rate at $N=500$ (\%)},
    xmin=0, xmax=100,
    ymin=0, ymax=60,
    xtick={0,20,40,60,80,100},
    ytick={0,10,20,30,40,50,60},
    grid=both,
    legend style={font=\small, at={(0.5,-0.35)}, anchor=north, legend columns=2},
]
\addplot+ coordinates {(0,0.00) (5,10.20) (10,12.98) (15,14.94) (20,19.80) (25,24.75) (30,28.42) (35,30.00) (40,33.60) (45,34.80) (50,41.00) (55,44.60) (60,45.36) (65,46.67) (70,47.86) (75,49.10) (80,49.80) (85,50.83) (90,52.00) (95,53.21) (100,54.20)};
\addlegendentry{\texttt{gpt-oss\_120b}}
\addplot+ coordinates {(0,0.00) (5,0.00) (10,5.52) (15,7.00) (20,11.96) (25,14.60) (30,15.88) (35,20.39) (40,22.44) (45,24.71) (50,26.80) (55,28.80) (60,31.28) (65,33.87) (70,35.66) (75,36.25) (80,40.12) (85,44.23) (90,46.20) (95,48.80) (100,51.80)};
\addlegendentry{\texttt{gpt-oss\_20b}}
\addplot+ coordinates {(0,0.00) (5,0.00) (10,0.00) (15,0.00) (20,0.00) (25,0.00) (30,0.00) (35,0.00) (40,0.00) (45,0.00) (50,0.00) (55,0.00) (60,0.00) (65,0.00) (70,0.00) (75,0.00) (80,0.00) (85,0.00) (90,0.22) (95,11.80) (100,27.00)};
\addlegendentry{\texttt{llama3.3\_70b}}
\end{axis}
\end{tikzpicture}
\caption{Quantile plot of $N=500$ match rates for three models. The curves reveal instability: some models spend a large fraction of runs near 0\%, with occasional high-coverage outliers.}
\label{fig:quantiles}
\end{figure}
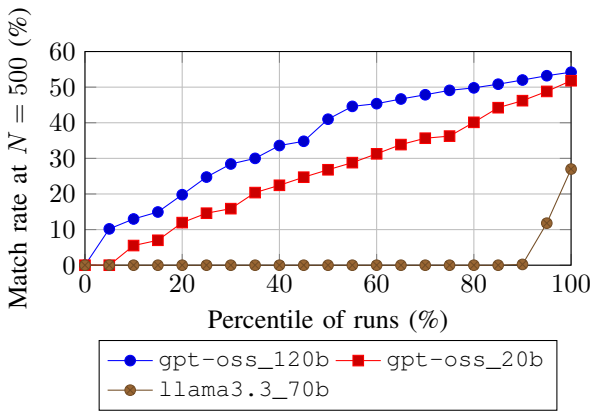

\section{Why state-of-the-art systems remain brittle on verbatim transcription}
It is natural to ask why modern models---including code-tuned LLMs---do not simply ``get better'' with scale on this task.
The issue is not that models cannot represent the data, but that the standard language-model training objective and decoding pipeline provide no guarantee of exact reproduction over long horizons.

At training time, autoregressive LLMs are optimized to minimize average token-level loss, not to guarantee that a particular long output is perfectly correct.
If the per-token probability of an error is small but nonzero, the probability of producing a completely error-free sequence of length $T$ decays roughly as $(1-\epsilon)^T$, which becomes tiny as $T$ grows.
This simple exponential effect helps explain why our experiment looks like a phase transition: at small $N$ the system often stays on-track, but beyond a length threshold the expected number of small slips becomes large enough that perfect transcription is rare.

At inference time, additional factors make exactness harder. Sampling-based decoding introduces variability; even greedy decoding can drift once an early token is wrong because future tokens are conditioned on the model's own previous output.
Instruction-tuning can also work against verbatim reproduction because the model is rewarded for being ``helpful'' and concise, which may bias it toward summarizing or restructuring rather than emitting long, repetitive sequences.
These tendencies overlap with broader reliability concerns discussed in the hallucination literature, where fluent output can deviate from strict ground truth and mitigation remains an open challenge \cite{huang2023hallucination}.

Many recent advances focus on \emph{input} length, such as extended context windows, position-embedding scaling, or retrieval augmentation.
Those methods can help a model \emph{access} the right information, but they do not automatically make the \emph{output} reliable at token-level fidelity.
Long-context benchmarks consistently report that effective use of long sequences remains brittle as context length and task complexity increase \cite{bai2023longbench,hsieh2024ruler,liu2023lost}.
Our benchmark is complementary: it stresses not only the ability to read long input, but also the ability to maintain output integrity over thousands of generated tokens.

Numbers further amplify this brittleness. Subword tokenizers can fragment numeric strings into multiple pieces, inflating sequence length and making small edits easy to introduce \cite{thawani2021numbers}.
Even when the underlying task is simple, numerical capabilities in transformer models often degrade out of distribution, such as when extrapolating to longer sequences or longer digit counts \cite{pal2021numeracy}.
Probing evidence also suggests that some LLMs internally represent numbers in digit-wise form, which helps explain digit-level error patterns rather than smooth numeric noise \cite{levy2024digit}.
From an architectural viewpoint, theoretical and empirical analysis connects failures in counting and copying to information ``over-squashing,'' where many earlier tokens have vanishing influence on the next-token prediction \cite{barbero2024glasses}.

Taken together, these points explain why even state-of-the-art models remain unreliable for integrity-critical verbatim transcription. The practical remedy is to treat the LLM as a planner and use deterministic tools for exact emission and verification, such as writing an external data file, producing a checksum, or using constrained decoding against a machine-checkable grammar.
This tool-backed approach aligns with evidence from code-security studies that LLM-generated code still requires careful validation and review \cite{pearce2021copilot,tihanyi2024secure,basic2024codesecurity}.

\section{Security implications for AI-assisted coding}
From a security perspective, the most important property in this benchmark is not that models make mistakes.
It is that the mistakes are easy to miss.
A generated file that \emph{looks} like correct code can embed a subtly corrupted dataset, and the surrounding logic can still execute without raising an exception.
This is a form of silent data corruption, except that it happens at the boundary between natural-language prompting and code artifacts.

Several security-relevant workflows resemble our benchmark.
Consider allowlists and denylists, where missing a single entry changes policy.
Consider cryptographic constants or protocol test vectors, where one incorrect digit can invalidate verification logic or defeat reproducibility.
Consider security baselines encoded as configuration arrays, where omissions reduce coverage without obvious breakage.
In these settings, an LLM that ``mostly copies'' is not a safe assistant unless the pipeline treats its output as untrusted and validates it.

The benchmark also highlights a subtle supply-chain risk.
In modern development, LLM output may enter repositories through copy-paste, code review, or automated tooling.
Reviewers often focus on control flow and API usage, not on verifying hundreds of numeric literals.
If the literal list is long, reviewers may not even attempt full verification, and diffs can be visually overwhelming.
That is precisely why a long-list transcription test is valuable: it forces an integrity failure that a normal review workflow is likely to overlook.

A defensible workflow therefore needs explicit integrity checks.
One approach is to avoid embedding large data entirely and load it from a versioned external file, so the model generates a loader rather than a literal transcription.
When embedding is unavoidable, the generated code should include machine-checkable assertions, such as verifying list length, checking a checksum of the concatenated literals, or reconstructing the data through a parsing step that is validated against a canonical representation.
The key principle is that correctness must be verified by deterministic tooling, not inferred from the plausibility of the generated text.

\section{Why this benchmark is a strong stress test, and how to strengthen it}
This task is a good stress test because it is simple to define, hard to cheat, and hard to solve by partial reasoning.
It also scales smoothly: increasing $N$ increases the required output volume and increases the duration over which state tracking must remain stable.
High-precision numbers increase the entropy per line and reduce redundancy, making errors both likely and measurable.

Future versions can strengthen the stress test by changing the structure of the data and the invariants the code must satisfy.
For example, near-duplicate literals that share long prefixes but differ in a few trailing digits can detect whether the model is collapsing suffixes.
Mixing signed numbers, scientific notation, and edge cases such as extremely small magnitudes can test format normalization and parsing behavior.
Embedding structured records, such as JSON objects with multiple numeric fields, can test whether state tracking generalizes beyond flat lists.
Introducing explicit cross-checks, such as requiring the model to output both the data and a hash computed over the canonical string representation, can separate truncation failures from rewriting failures because any change in formatting will break the hash.

The validator can also be strengthened.
Substring matching is intentionally strict and intentionally simple, but a security-oriented evaluation may additionally want to parse the generated code, extract the list programmatically, and compare it to a canonical representation.
That would allow the analysis to distinguish missing values from reformatted values, quantify duplication, and localize whether missing values cluster at the tail (suggesting truncation) or appear throughout (suggesting drift).

\section{Limitations}
The conclusions in this paper are bounded by the validation strategy and the available artifacts.
Because we evaluate exact-string inclusion, numerically equivalent but reformatted outputs count as failures.
Because we do not analyze raw generated outputs, we cannot directly attribute failures to truncation, refusal, or format drift, even when the aggregate patterns strongly suggest those mechanisms.
Finally, the task family is narrow by design; it isolates data transcription, and it does not measure algorithmic reasoning or broader code quality.

\section{Conclusion}
The experiments show that verbatim data transcription is a fragile capability in current LLM-based code generation.
Some models can reliably embed 100 high-precision numbers into Python code, yet none of the evaluated models achieve a perfect transcription at 300 or 500 numbers in the provided runs.
Long outputs reveal two problems at once: a capacity-limited ceiling that prevents complete transcription, and state-tracking derailments that produce outputs containing none of the expected data.
For security-sensitive pipelines, the lesson is straightforward: treat LLM outputs as untrusted, and verify integrity with deterministic tooling.
For researchers and evaluators, the benchmark offers a compact stress test that complements algorithmic code-generation metrics by measuring something that software engineers often need and often assume: accurate copying.
\section{AI Usage Acknowledgement} Chatgpt 5.1 is used to refine the language of this paper and latex edit and graph generation.
\bibliographystyle{abbrv}
\bibliography{name}



\end{document}